\documentclass[
reprint,  
dcolumn, 
superscriptaddress,
 amsmath,amssymb,
 aps,
]{revtex4-2}

\usepackage{graphicx}
\usepackage{dcolumn}
\usepackage{bm}
\usepackage[bottom]{footmisc}
\usepackage{xcolor}

\usepackage{appendix}

\begin{document}

\title{Observations of Low-Energy-electrons production and experimental characterization of test mass charging process in the LISA gravitational reference sensor with the BART (Beam Assisted Radiation Test) experiment)}

\title{Observations of Low-Energy-Electron Production and Experimental Characterization of the Test-Mass Charging Process in the LISA Gravitational Reference Sensor with the BART Experiment}

\author{Francesco Dimiccoli}
\email{francesco.dimiccoli@unitn.it}
\thanks{ORCID: 0000-0002-8766-9625}
\affiliation{Department of Physics, University of Trento, Via Sommarive 14, I-38123 Trento, Italy}
\affiliation{INFN-TIFPA, Via Sommarive 14, I-38123 Trento, Italy}

\author{Francesco Venturelli}
\email{francesco.venturelli@unitn.it}
\thanks{ORCID: 0009-0008-2604-1907}
\affiliation{Department of Physics, University of Trento, Via Sommarive 14, I-38123 Trento, Italy}

\author{Valerio Ferroni}
\email{valerio.ferroni@unitn.it}
\thanks{ORCID: 0000-0002-2260-6658}
\affiliation{Department of Physics, University of Trento, Via Sommarive 14, I-38123 Trento, Italy}
\affiliation{INFN-TIFPA, Via Sommarive 14, I-38123 Trento, Italy}

\author{Antonella Cavalleri}
\email{antonella.cavalleri@unitn.it}
\affiliation{Department of Physics, University of Trento, Via Sommarive 14, I-38123 Trento, Italy}

\author{Teodoro Klaser}
\email{teo.klaser@outlook.it}
\thanks{ORCID: 0000-0000-0000-0004}
\affiliation{Department of Physics, University of Trento, Via Sommarive 14, I-38123 Trento, Italy}

\author{Matteo Tomasi}
\email{matteo.tomasi-1@unitn.it}
\thanks{ORCID: 0009-0007-4529-1904}
\affiliation{Department of Physics, University of Trento, Via Sommarive 14, I-38123 Trento, Italy}
\affiliation{Department of Industrial Engineering, University of Trento, I-38123 Trento, Italy}

\author{Davide Dal Bosco}
\email{davide.dalbosco-1@unitn.it}
\thanks{ORCID: 0009-0009-4054-3855}
\affiliation{Department of Physics, University of Trento, Via Sommarive 14, I-38123 Trento, Italy}
\affiliation{INFN-TIFPA, Via Sommarive 14, I-38123 Trento, Italy}

\author{Enrico Verroi}
\email{enrico.verroi@tifpa.infn.it}
\thanks{ORCID: 0000-0003-0957-4505}
\affiliation{INFN-TIFPA, Via Sommarive 14, I-38123 Trento, Italy}

\author{Rita Dolesi}
\email{rita.dolesi@unitn.it}
\thanks{ORCID: 0000-0002-4191-7558}
\affiliation{Department of Physics, University of Trento, Via Sommarive 14, I-38123 Trento, Italy}
\affiliation{INFN-TIFPA, Via Sommarive 14, I-38123 Trento, Italy}

\author{William J.~Weber}
\email{williamjoseph.weber@unitn.it}
\thanks{ORCID: 0000-0003-1536-2410}
\affiliation{Department of Physics, University of Trento, Via Sommarive 14, I-38123 Trento, Italy}
\affiliation{INFN-TIFPA, Via Sommarive 14, I-38123 Trento, Italy}
```

\date{\today}

\begin{abstract}
The Laser Interferometer Space Antenna (LISA) is a space-based gravitational-wave observatory that uses free-falling test masses as inertial references to detect milliHertz-frequency signals. Interactions between these test masses and galactic or solar energetic particles cause charge buildup and Coulomb forces, a major source of acceleration noise that must be accurately modeled. LISA Pathfinder measurements showed that the Poissonian test mass charging noise, an important entry in the noise budget, was considerably larger than that in pre-launch simulations, indicating missing physical processes in early models. Emission of low-energy secondary electrons (LEE) from test mass and housing surfaces has been proposed as a key mechanism affecting charging and the sensor’s electrostatic response.

We report a particle accelerator-based experiment that directly tests the LEE hypothesis by measuring proton-induced test mass charging in a LISA-like Gravitational Reference Sensor geometry as a function of the test mass electrostatic potential. A flight-representative test mass and the capacitive sensor electrode housing surrounding it are exposed in vacuum to monoenergetic 70–230 M$e$V proton beams, while test mass charging currents are measured with femtoampere sensitivity. By varying the test mass potential by volts during irradiation, we gain sensitivity to eV-scale charge carriers whose transport and collection depend on these fields. The measurements yield stable, reproducible picoampere-level charging currents and a clear dependence of the net charging rate on test mass potential. This behavior directly supports the role of low-energy secondary electrons in test mass charging and is consistent with the LEE-based interpretation of LISA Pathfinder data.

These are the first controlled laboratory measurements of low-energy electron effects in test mass charging for a LISA-like sensor and provide a benchmark for validating charging models and charge-management strategies for LISA and future space-based precision interferometry missions.

\end{abstract}

\maketitle


\section{Introduction}\label{sec_intro}

In 2015, gravitational-wave physics entered a new era when the Laser Interferometer Gravitational-Wave Observatory (LIGO) detected gravitational waves originating from a binary black hole merger, providing the first direct confirmation of Einstein’s General Relativity \cite{PhysRevLett.116.061102}. In the same year, the European Space Agency (ESA) launched the \textit{LISA Pathfinder} (LPF) mission, which successfully demonstrated that a test mass (TM) could maintain geodesic motion with residual acceleration noise below $10^{-14}~\mathrm{m/s^2/\sqrt{Hz}}$ at mHz frequencies, thereby paving the way for a space-based gravitational wave observatory.

Approved by ESA in 2017 and formally adopted in 2024, the \textit{Laser Interferometer Space Antenna} (LISA) mission is scheduled for launch in 2035. LISA will consist of three spacecraft in a triangular constellation with 2.5 million km arms, each hosting two free-falling TMs made of a gold–platinum alloy. These TMs act as inertial references for tracing the gravitational wave tidal deformation, which is then measured by laser interferometry. Each TM is enclosed, without any mechanical or electrical connection, within a molybdenum gold-plated electrode housing (EH), which provides capacitive sensing and actuation while maintaining mechanical and electrical isolation. However, exposure to galactic and solar cosmic rays (CR) leads to continuous TM charging, which couples with stray electric fields in the EH to produce spurious Coulomb forces. In-band acceleration noise arises due to both stray fields and TM charging fluctuating in time \citep{grimani,Shaul:2005tz}. As taught by LPF, to mitigate these effects, LISA will measure and electrostatically compensate in flight to a few mV the average s[tray fields\citep{PRLStray,PhysRevLett.118.171101}, and incorporate a dedicated \textit{Charge Management System} (CMS) that employs ultraviolet (UV) illumination to actively discharge the TMs via the photoelectric effect, thereby maintaining its charge within a few mV to neutral condition. However, CMS does not suppress the noise arising from stochastic charge fluctuations associated with the Poissonian nature of the charging process\footnote{In fact, photo-electron currents contribute to charge fluctuations.}. A theoretical characterization of TM charging through Monte Carlo simulations remains essential, both for determining the net charging rate under the wide range of interplanetary conditions and for quantifying the corresponding charge fluctuations.

To this purpose, extensive Monte Carlo studies of the charging effect have been performed since the LPF development phase, using both \textsc{FLUKA} \citep{Battistoni2015FLUKA,Ahdida2022FLUKA} and \textsc{GEANT4} \citep{Geant4,1610988,ALLISON2016186}. In-flight LPF measurements revealed a positive net charging rate of $\sim$ 25 e/s, in line with the expectations from the simulations of the time, while \citep{PhysRevLett.118.171101} measured a TM charge noise with a 1/f$^2$ noise power spectrum, consistent with Poissonian charging events with an effective event rate between 1000 and 1400 $s^{-1}$ roughly 4-5 times larger than the pre-flight predictions \citep{Wass_2005}\citep{PhysRevD.107.022010}\citep{grim15}. Moreover, the TM charging was found to decrease linearly with increasing TM potential, falling to zero at an equilibrium potential of 900 and 951 mV, respectively, for the two TMs. 
Those discrepancies were attributed to the contribution of a quasi-stationary cloud of order $\sim$ $e$V secondary electrons \citep{ARAUJO2005451,Nostro,PhysRevD.107.062007_Charging2023}which populates the gap between the TM and the EH.
This interpretation proposed that CR-induced charging is self-limited by the emission of electrons with eV-level kinetic energies by the gold surfaces of LISA TM and EH in the vacuum gap between them. Under the effect of voltage biases, these electrons can migrate to counterbalance the net TM charging, potentially explaining the TM voltage saturation effect observed by LPF.

In 2023, an ESA-funded project conducted by the University of Trento, the University of Urbino, and OHB Italy (ESA ITT Ref.: AO/1-10081/20/NL/CRS, \textit{``Test Mass Charging Toolkit and LPF Lessons Learned''}) focused on characterizing the emission of Low Energy Electrons (LEE) and incorporating their contribution into a new Monte Carlo simulation framework based on the GEANT4 engine.  
From a theoretical standpoint, the impact of LEE emission from the TM and EH surfaces under CR irradiation was analyzed in a joint effort between the University of Trento and the Trento ECT* group.  
Secondary LEE are generated primarily through \textit{kinetic emission}, namely emission predominantly originating from electron–electron interactions in which ionization electrons lose nearly all of their energy via inelastic scattering processes \citep{Nostro}.  
The LEE yield from gold surfaces was constrained by combining \textit{ab initio} theoretical calculations with the limited experimental data available from electron backscattering measurements \citep{Cimino}. No data instead are available to constrain emission subsequent to hadron irradiation. This encouraged the development and validation of a \textsc{GEANT4}-based toolkit (the Test Mass Charging Toolkit, TMCTK) designed to provide an end-to-end simulation of the effect of TM charging on the LISA sensitivity over a broad range of space-weather conditions, including both long-term CR flux variations and transient events. The toolkit incorporates the simulation of LEE using GEANT4 simulation packages
specifically developed for the description of very-low energy electromagnetic interactions in gold (GEANT4-DNA) down to energies of a few eV, thereby yielding improved predictions of TM charging and its associated contribution to LISA’s acceleration noise \citep{toolkit}.  Nevertheless, the charging-noise predictions produced by TMCTK, although in closer agreement with LISA Pathfinder (LPF) measurements than earlier GEANT4-based simulation models, remain underestimated. This discrepancy motivates further investigations of LEE emission and of the TM charging process in general, with particular emphasis on experimental studies employing configurations more representative of actual flight conditions in the interplanetary environment, where accelerated hadrons of cosmic and solar origin interact with the spacecraft. Such experiments will enable controlled tests of GEANT4 predictions and support the refinement of its models for the production and transport of extremely low-energy particles.

In this work, we address this objective by presenting the results of the Beam Assisted Radiation Test (BART) experiment, specifically conceived to investigate, for the first time in a controlled laboratory environment, the TM charging process. The primary aim of BART is to provide experimental confirmation of LEE emission induced by the exposure of a LISA-representative geometry to energetic hadrons (protons), quantifying their contribution and characterizing their emission spectrum.

 \begin{figure}[t]
\centering
\includegraphics[width=0.45\textwidth]{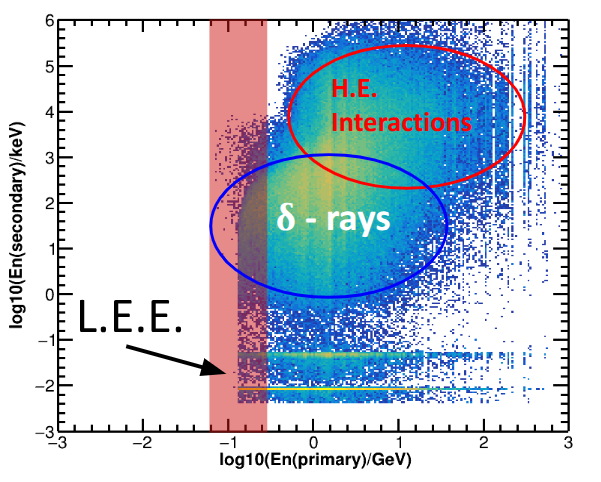}
\caption{Energy distribution of secondary electron produced in the gap between TM and EH from a simulated flux of cosmic primary protons hitting a LISA representative geometry using the TMCTK simulation tool (\cite{toolkit}), as a function of the primary proton energy. The insert shows the projection of the 50-230 M$e$V slice (red band), representative of the energy range accessible by the Proton Therapy cyclotron (see Section \ref{sec_apparatus}). In the plot, the cotribution coming from secondary electrons generated by high energy (H.E.) interaction such as pair production (red circle) and $\delta$-ray emission (blue circle) are indicated. }\label{fig:lee}
\end{figure}

\subsection{Observing TM charging on-ground: The BART experiment}

The idea of the BART experiment is to use a controlled energetic proton source to measure the TM charging phenomenon and the emission of LEE with an apparatus representative of the LISA in-flight geometry and material configuration, exploring an energy range accessible by mid-scale research particle accelerators like a cyclotron, up to 250 M$e$V. \textcolor{blue}{Although a beam-like irradiation differs from the isotropic cosmic-ray environment expected in flight, it provides a controlled benchmark of the relevant charging and low-energy electron emission processes.} An idea of the level representativeness of this test with respect to the on-flight TM charging phenomenon can be obtained from the knowledge of the contributions to the global charging rate and charging noise of primary CR of different energies, that can be studied thanks to the TMCTK.

The TM net charging rate ($\lambda_{NET}$) is mainly due to low-energy protons stopping in the TM and from secondary particles generated by higher-energy proton interactions. TMCTK simulations under stationary solar modulation show that $\lambda_{NET}$ is mainly due to primary protons in the 70-1000 M$e$V range, stopping in the TM. Higher-energy protons are not effectively stopped and thus do not directly charge the TM. Protons below 50–100 M$e$V are stopped in surrounding materials, dominated by the EH, and also do not contribute. The simulations indicate a significant contribution also from secondary electrons to TM charging \citep{11114657}. At sub-G$e$V primary proton energies, their contribution is subdominant with respect to the contribution of the stopping protons. For example, for primary proton energies of 100–300 M$e$V, their net charging contribution is negative, reducing the net rate by 10–20\%, because electrons entering the TM from EH emission outweigh electrons emitted from the TM. At energies $\geq 2$ GeV, where protons traverse the TM, secondary electrons dominate the charging, and their net contribution is found from simulations to be positive \citep{toolkit}. 

 \begin{figure}[t]
\centering
\includegraphics[width=0.45\textwidth]{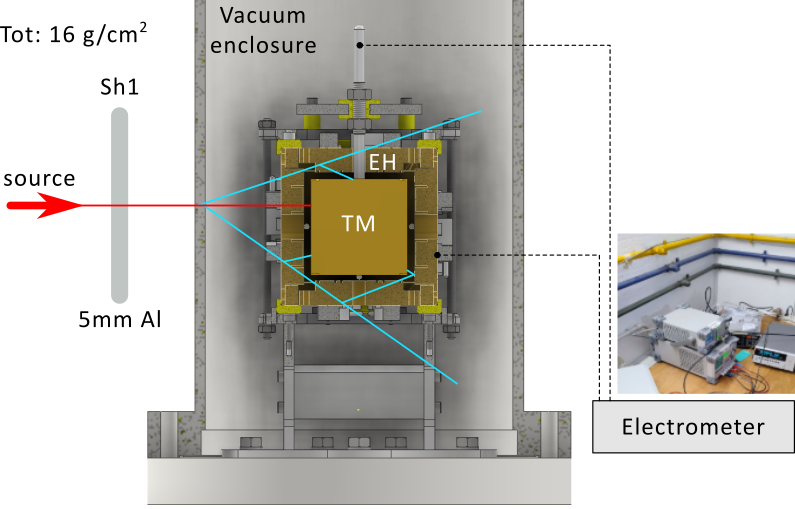}
\caption{Scheme of the BART experiment concept: The TM is kept electrically isolated from the EH by Teflon spacers (shown in light yellow in the scheme) and charged up by a proton (red track), creating secondary particles (blue tracks). The charging is measured as a discharge current by an electrometer.}\label{fig:scheme}
\end{figure}

Figure \ref{fig:lee} shows the spectrum of secondary electrons produced in the gap between  EH and TM as a function of the energy of the primary protons, of a cosmic proton flux typical of the solar minimum condition, using the TMCTK tool, as a function of the energy of the primary protons. The distribution shows three main contributions: the majority of electrons show energy from 100 $e$V to 1 M$e$V and are produced by delta ray emission from high energy charged particles. Electrons of higher energies, above 1 M$e$V can be produced by interactions such pair production or pion decay. Finally, the simulation shows that significant production of LEE below 100 $e$V exists across the majority of the primary proton spectrum as a consequence of \textit{kinetic emission}. The red band in Figure \ref{fig:lee} corresponds to the 50-230 M$e$V energy range, accessible by BART. Therefore we can study the threshold of the TM charging activation and produce a detectable amount of LEE, which can be experimentally probed by varying the potential difference between the TM and EH. Figure \ref{fig:scheme} shows a scheme of the experiment, the configuration of which is described in detail in Section \ref{sec_apparatus}.

\section{Experimental Apparatus} \label{sec_apparatus}

 \begin{figure}[t]
\centering
\includegraphics[width=0.35\textwidth]{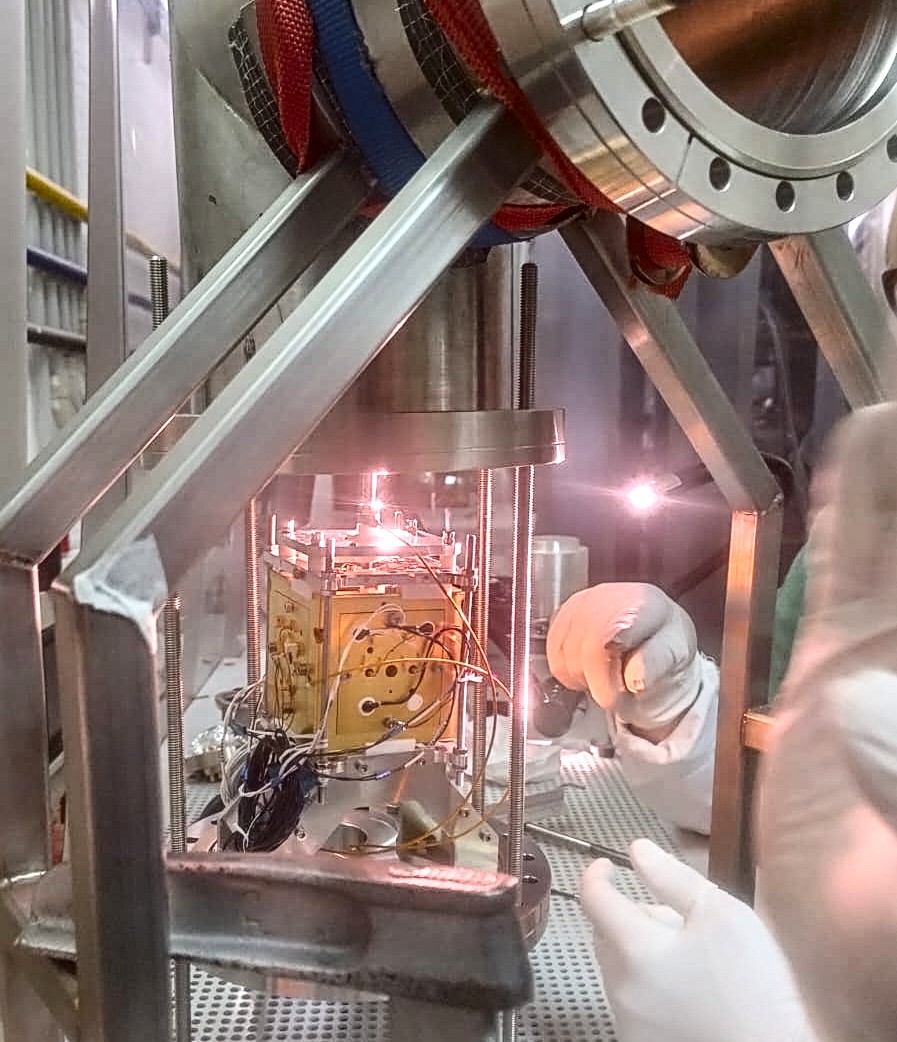}\vspace{0.5cm}
\includegraphics[width=0.4\textwidth]{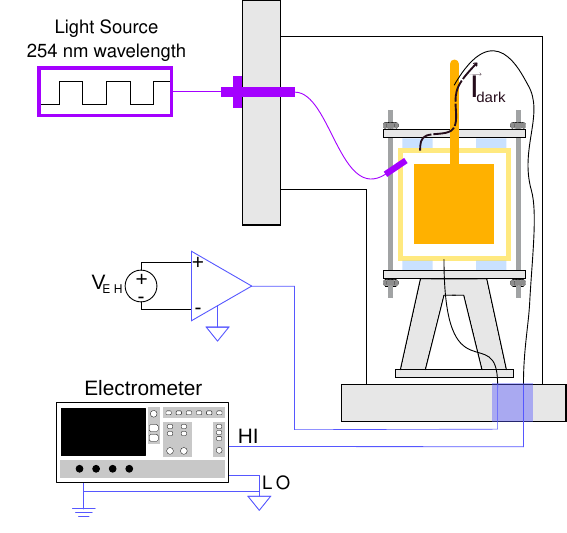}
\caption{Top: Integration of the EM of the LPF GRS inside the BART vacuum chamber. Bottom: Scheme of the electrical and optical (used in performance testing) configuration of the experiment. The dashed arrow  shows one of the dark current paths through the Teflon insulators.}\label{fig:scheme2}
\end{figure}

The Proton irradiation for the experiment is provided by the Trento Proton-Therapy (PT) facility, which delivers a beam with energies between 70 and 230\,M$e$V, described in more detail in Section \ref{pt_facil}. This beam is directed to an experimental setup designed to provide the best possible representativeness of the LISA GRS sensor and its flight conditions, with the possibility to monitor in the most precise way the charge transfer processes induced by the interaction of energetic protons.

At the core of the BART setup lies a gold-coated copper test mass, that replicates the exact size, geometry, and surface finish of the LISA flight model, except for a conductive shaft on one face, that we also use as an electrical contact to the TM. The TM is housed within a molybdenum electrode housing (EH), originally developed as an engineering model for the LISA Pathfinder mission. The EH is configured as a cubic enclosure with wall thicknesses ranging between 0.9 and 1.1 cm, and it incorporates gold–sapphire \footnote{The electrodes employed in this study differ from those used in the LPF flight model, which were molybdenum electrodes coated with gold (Au) isolated by sapphire spacers \citep{Dolesi_2003}. This discrepancy influences only the material budget in the vicinity of the TM and is not expected to affect any other aspects relevant to the present analysis.} electrodes embedded within 3 mm-deep grooves along its internal surfaces. These electrodes are employed for capacitive sensing of the TM position within the GRS, for TM AC polarization (“injection” electrodes), and for TM actuation. The electrode layout is not uniform across all faces of the sensor, yielding a face-dependent variation in the effective thickness of the EH to traversing particles. Specifically, the EH hosts two smaller electrodes on the faces perpendicular to the science measurement axis (hereafter denoted as “face X”) and three slightly larger electrodes on the faces parallel to that axis (“face Y”), leading to a reduced average grammage on the latter. The resulting difference, averaged over the full area of an EH face, is estimated to be (1 $\pm$ 0.2) g/cm$^2$.

The EH is mounted on a DN160CF flange via a mechanically rigid but electrically insulated support structure (see Figure \ref{fig:scheme2}). The mechanical support equipment is necessary for holding both EH and TM inside the vacuum enclosure electrically isolated from each other and from the vacuum chamber. The EH is clamped into an aluminum structure, attached to the base of the vacuum chamber. Insulator elements made in PTFE, shown in light blue in Figure \ref{fig:scheme2}, are placed in between the clamps and the EH, allowing for electrical isolation between the two. 
The TM shaft goes through the face Z hole of EH for the Grabbing, Positioning and Release Mechanism (GPRM) \citep{tomasi2023preliminary}, and it is suspended on top of the EH clamp. Teflon rings with resistances in the tens of T$\Omega$, center the TM shaft. These high-resistance supports ensure minimal leakage currents which were measured in the preliminary tests of the facility (see Section \ref{sec_comm}).

\begin{figure}[h]
\centering
\includegraphics[width=0.25\textwidth]{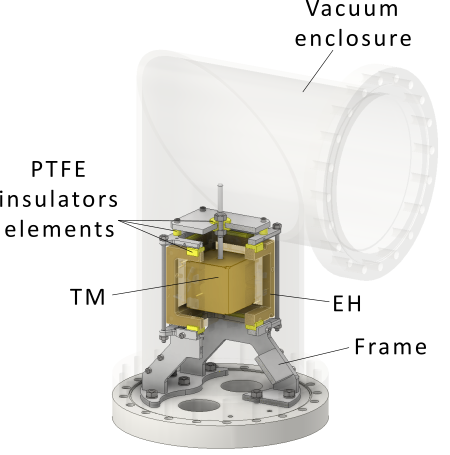}
\hfill
\includegraphics[width=0.2\textwidth]{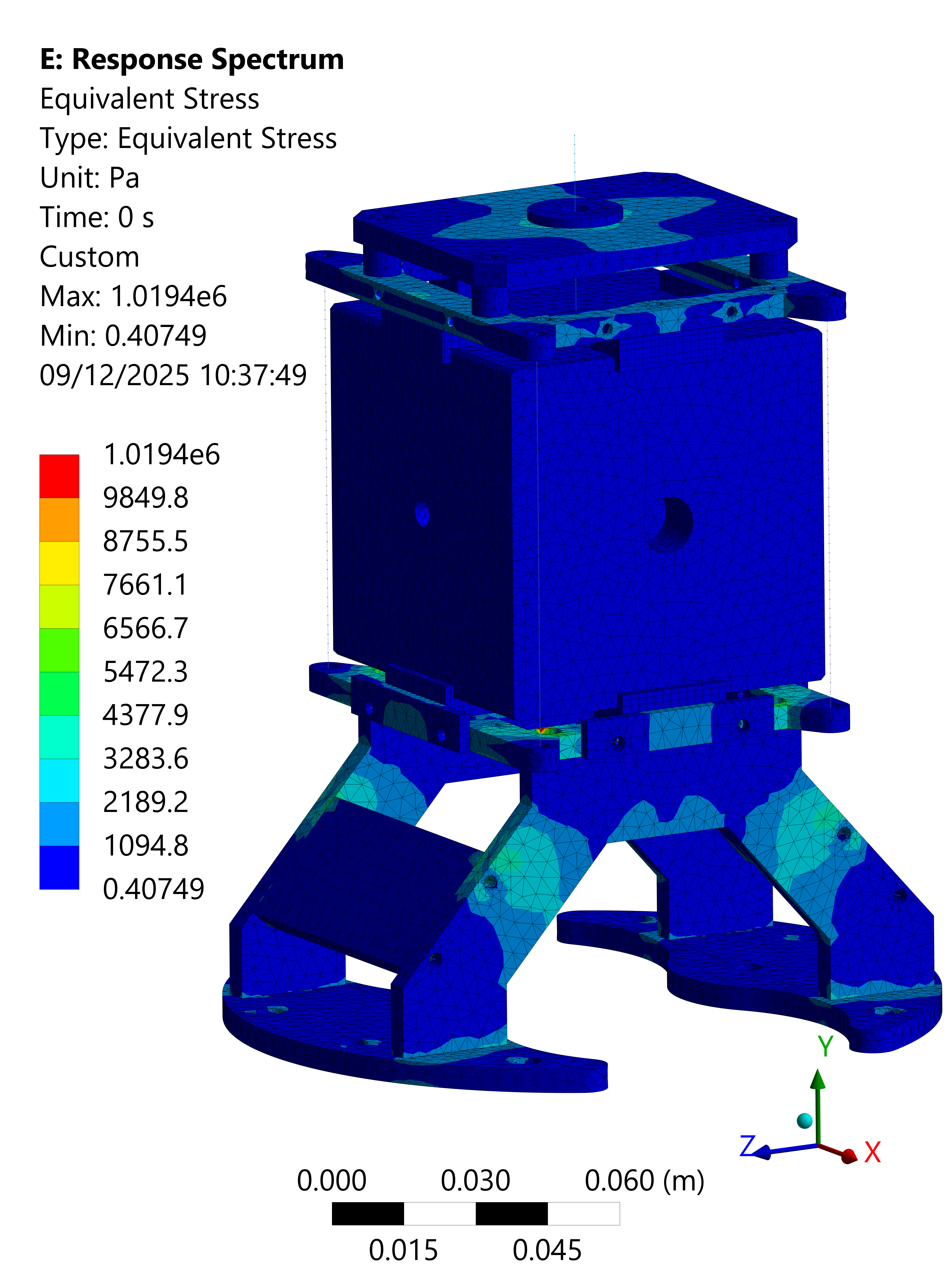}
\caption{CAD model of the mechanical support equipment used for the tests (left). The aluminum structure of the mechanical support equipment and the EH frame is sectioned, and EH's faces are hided to show the TM. Finite element results for the response spectrum analysis (right), evaluating handling and transportation loads.}
\label{fig:mgse_cad}    
\end{figure}

Mounting the basement of the structure directly to the bottom flange (Figure \ref{fig:mgse_cad}), gives enough clearance for the electrical cables and fiber optics inside. This also, enables an easy integration of the setup: assembling the setup on the vacuum chamber basement, and dropping the camber onto it at the end of the integration process.\\

The structural assembly accommodating the EH and TM was validated for resistance to transportation loads by means of a finite element analysis conducted in ANSYS 2025R2 (Figure \ref{fig:mgse_cad}). A response spectrum analysis, using the acceleration load spectra specified in \cite{ISO_13355}, confirmed that transportation-induced stresses do not compromise the integrity of the structure and yield negligible deformations.

The TM–EH assembly is enclosed in a 3.5 mm thick stainless-steel vacuum chamber with a diameter of 15\,cm and a height of 40\,cm. The chamber maintains a pressure below $10^{-6}\,\text{mbar}$ with a $30\;\text{l/s}$ turbo pump, so that the mean free path of electrons is much larger than the characteristic length of the experimental apparatus.  To reach a total effective material budget of approximately 13\, g/cm$^2$, comparable to in-flight conditions, an additive 5 mm wide Al shield is placed between the chamber and the beam source. The distance between the shield and the vacuum chamber can be adjusted to obtain a more uniform spread of the beam on the EH illuminated face, via multiple-scattering driven de-focusing of the beam.

The net sum of the electric charges entering/exiting the TM is monitored using a precision electrometer with femtoampere-level sensitivity. The system achieves a current resolution of approximately 10\,fA with a 10\,mHz modulation. This precision was evaluated via measurements of UV-induced photoemission currents as described in Section \ref{sec_comm}, confirming its capability to resolve the pA-level current expected by the TM charging process during the experiment.

 The electrostatic potential of the EH can be adjusted within the interval from $-10$\,V to $+10$\,V with respect to the TM, grounded through the amperometer, thereby enabling precise control of the electric field established in the TM–EH gap. The bias voltage is supplied by a generator and delivered via a dedicated front-end electronic circuit based on AD797 amplifiers \citep{AD797_datasheet}. 

The experimental configuration described above is designed to reproduce as closely as possible the in-flight LISA environment and operational conditions, while ensuring high precision in the measurement of TM charging currents in different conditions of voltage bias. Structural discrepancies—such as differences in chamber geometry and supporting hardware—are not expected to introduce significant systematic uncertainties, given that the primary objective of the experiment is to investigate charge transport and deposition processes.

\subsection{The Trento Proton Therapy beam facility} \label{pt_facil}

The PT beam facility is operated by the \textbf{INFN-TIFPA (Trento Institute for Fundamental Physics and Applications)} (\url{https://www.tifpa.infn.it/}) and is 
located within the \textbf{Trento Proton Therapy Centre} (Azienda Provinciale per i Servizi Sanitari, Trento, Italy). 
The facility hosts a commercial 230~M$e$V IBA \textit{Proteus~235} cyclotron used for both clinical proton therapy and 
scientific research. Protons are extracted at a fixed energy of 230~M$e$V and transported through an energy-selection system 
comprising a graphite degrader, a magnetic analyzer, andcollimation slits, providing variable beam energies in the 
range 70--230~M$e$V. The energy spread is typically below 1\%, and energy switching can be achieved in less than one second~\citep{DiRuzza2020}.
 \begin{figure}[t]
\centering
\includegraphics[width=0.48\textwidth]{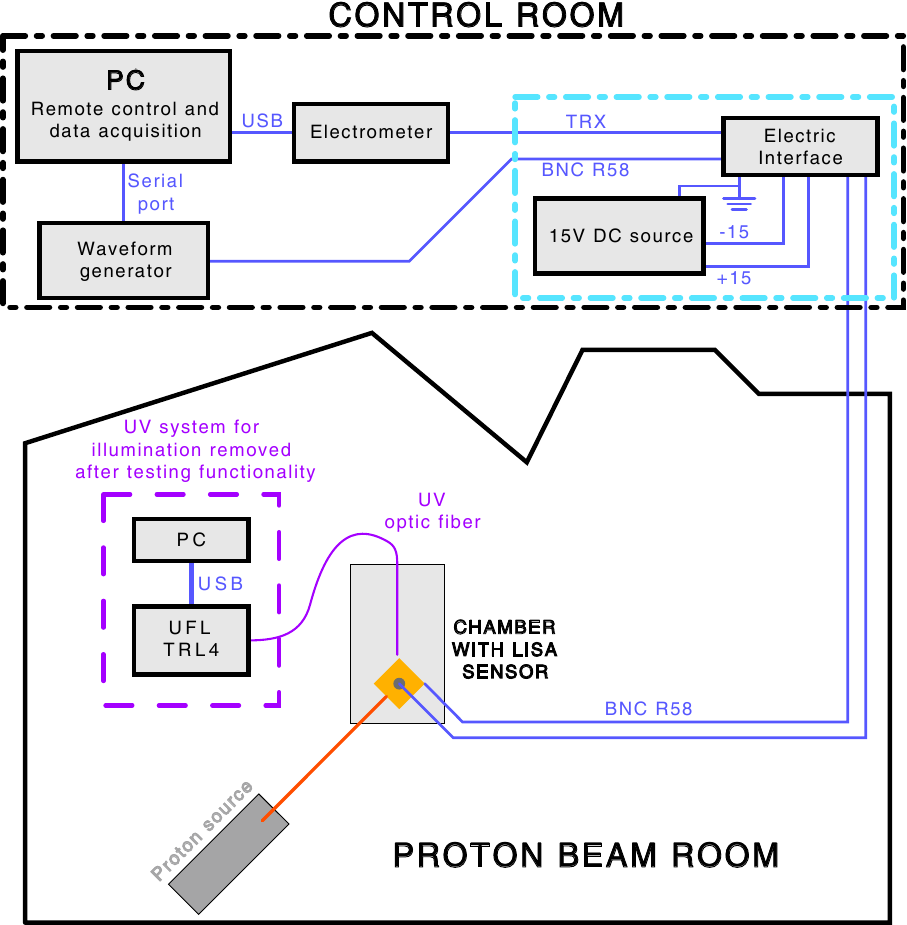}
\caption{Scheme of the configuration of the INFN-TIFPA PT experimental room for the BART experiment. }\label{fig:exp_room}
\end{figure}
A dedicated experimental area is connected to the accelerator through two horizontal beamlines: 
a ``biological'' line optimized for radiobiology and dosimetry studies, and a ``physics'' line designed for detector irradiation and instrument testing. 
The physics beamline allows both in-air and in-vacuum irradiation of 
experimental setups. The beam spot size and flux are adjustable, emitting fluences between 
$10^{6}$ and $10^{13}$~p\,cm$^{-2}$ per exposure with an intensity between 10$^6$ and 10$^9$ p/s. According to TMCTK simulations, rates above 10$^7$ p/s assured currents from LEE of the order of pA, resolvable by the BART electrometer.  The control of the beam energy through graphite degrader causes an increase of the beam cross area towards lower energies as a result of multiple scattering. The FWHM of the beam profile is measured to range from 6.8 mm at maximum nominal energy (228 M$e$V) to 16.2 mm at 70 M$e$V \citep{Benedetto2019}.
Beam intensity is monitored using calibrated transmission ionization chambers, 
while the absolute proton flux is determined through cross-checked Faraday-cup and secondary-emission monitors. 
Beam energy spectra and spatial profiles are routinely bench-marked with \textsc{Geant4} simulations and 
radio-chromic film measurements~\citep{Geant4,PubMed30824157}.
The irradiation geometry is fixed horizontally, with the beam entering the experimental room at approximately 
1.3~m above the floor level. The facility includes a controlled-access area with 1~m-thick concrete shielding and 
remotely operated positioning systems. 
 Figure \ref{fig:exp_room} shows the experimental room configuration set up for the BART experiment. Thanks to the built-in wiring of the experimental room, it was possible to operate the front-end electronics and the measurement apparatus remotely, from a dedicated control-room where also the beam control was managed, allowing to minimize noise on the read-out apparatus.

\subsection{TMCTK simulation tool for BART} \label{sec_sim}

 \begin{figure}[t]
\centering
\includegraphics[width=0.35\textwidth]{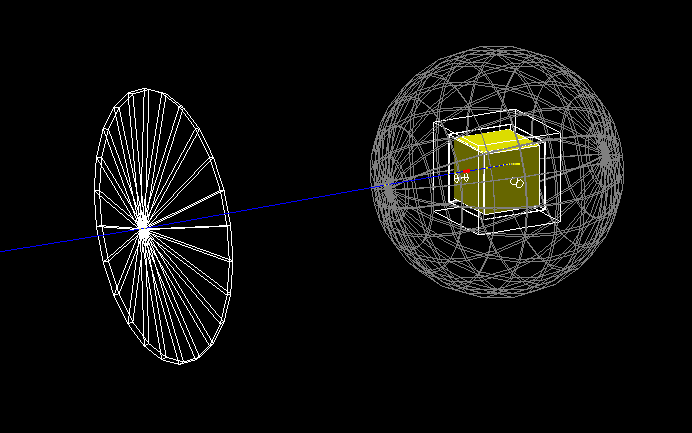}\\
\includegraphics[width=0.35\textwidth]{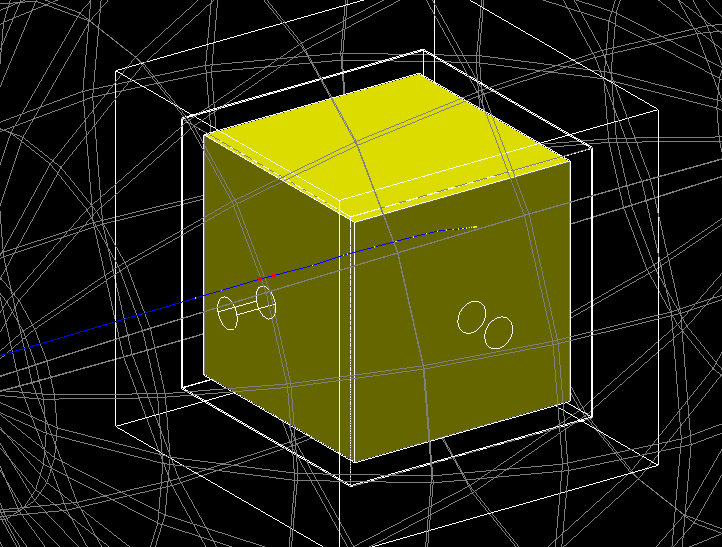}
\caption{Graphical rendering of the BART apparatus simulation. In blue a primary proton track stopping in the TM is shown. For display purposes, EH, vacuum chamber (grey spherical shell) and Al shield are shown in wire-frame mode. }\label{fig:geant}
\end{figure}
The TMCTK toolkit consists of a GEANT4 based module simulating the propagation of the cosmic particles and the secondary particles in the materials of LISA and of a COMSOL-based \citep{multiphysics1998introduction}  Finite Element Model (FEM) describing the electrostatic effects on the secondary electron trajectories in the gap between TM and EH.
The extremely simplified and symmetric geometry implemented in the TMCTK, suitable for the simulation of the effects of isotropic CR flux, was adapted to better represent the asymmetric beam-like irradiation of BART. A more complex geometry was implemented, substituting the concentrical Al shells of the standard TMCTK with a series of volumes namely describing the BART vacuum chamber, represented by a 3 mm wide stainless steel spherical shell, the structure of the EH, the TM model and the geometry of the proton beam.  The EH is modeled by a  Molybdenum (Mo) cubic shell, internally coated by 100 nm thick gold cubic shell, representing the EH internal gold coating. The EH features a 6 mm diameter circular hole in the middle of every cube face, modeling the interferometry laser windows\footnote{Only one of these holes actually serves as laser window in LISA, the others assure the required symmetry in mass distribution and vacuum conductance}. These holes are relevant because of the beamed flux of the incoming radiation. The width of the Mo volume is chosen as a result of a weighted average of the real material distribution of the EH, which includes less dense materials (i.e. sapphire electrodes surrounded by mm-wide vacuum gaps). To account for the known difference in material budget between faces X and Y of the EH (see Section \ref{sec_apparatus}), two different simulations were performed with Mo volume widths of 7.75 mm and 9.25 mm. This width difference compensates for the expected difference in average grammage from the distinct electrode configurations of the two faces.  
The modeling of the TM is adapted from that implemented in TMCTK, with the material of the transition-metal bulk substituted from gold to copper. As in TMCTK, the TM is immediately surrounded by a gold cubic shell modeling its gold coating.
Finally, the proton irradiation is modeled generating a beam of protons characterized by a Gaussian intensity profile, with a full width at half maximum varying from 6.8 to 16 mm depending on beam energy, on the basis of the specifications of the PT facility, described in Section \ref{pt_facil}. An 1\% smearing in beam energy is added to simulate the nominal energy spread of the PT beam. The additional Al shielding described in Section \ref{sec_apparatus} is also included in the simulation. A graphical rendering of the simulation geometry is represented in Figure \ref{fig:geant}, that also shows the trajectory of a proton hitting the side of the GRS facing the beam in a typical experimental configuration.

The final output of the TMCTK GEANT4 module is the net signed deposited charge on the TM in units of elementary charge per primary proton that allows for the calculation of the net charging on the TM. In addition, GEANT4 yields the position and the momentum of each LEE contributing. The electrostatic module of the TMCTK allows for tracing the trajectories of these electrons across the lines of the electric fields in the gap between TM and EH and for computing the actual charge LEE deposited on the TM. The model allows for the computation of the electric field lines and the tracing of charged particles in the TM-EH gap. Only LEE with energy $E_e$ large enough to overcome the energy barrier from material to vacuum ($W_F$) are considered, according to the condition\footnote{This condition is needed because our GEANT4 implementation does not include the contribution of the work function of the material in the propagation of the electrons.} $E_e\cos^2\theta>W_F$ \citep{Nostro}. The GEANT4 and the electrostatic model have offered the capability to fully simulate the BART experiments providing, at the beginning, important insights for the design of the experiment and then the benchmarks against which we compare to test the limits of the LEE modelization.
 
 \begin{figure}[t]
    \includegraphics[width=0.45\textwidth]{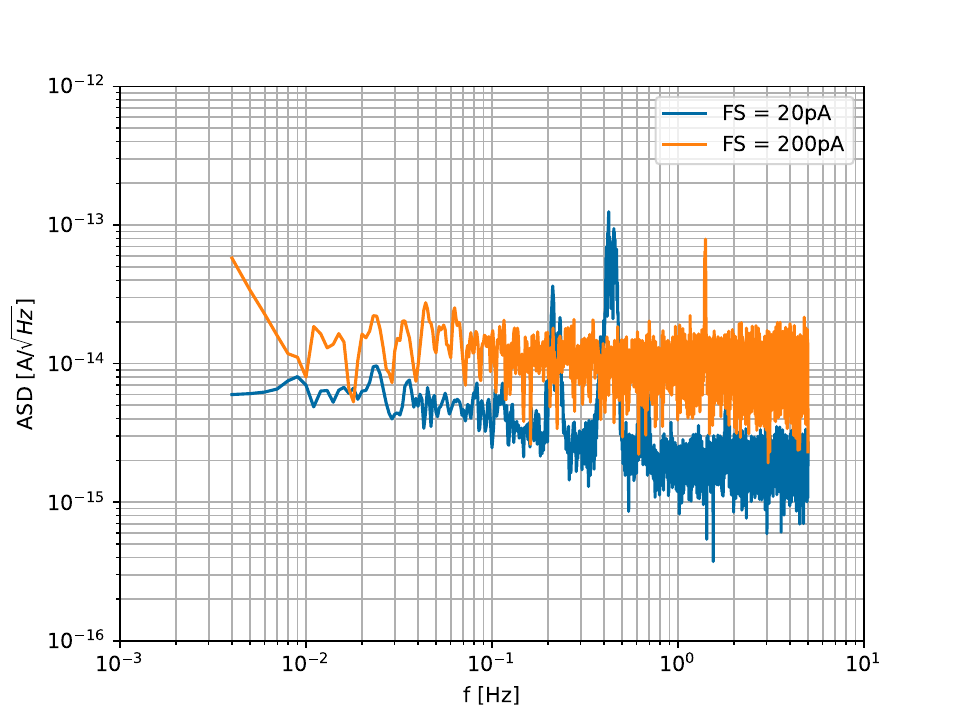}
    \caption{Noise spectrum of the BART electrometer with the two full scales used. It is almost white as expected from a digitized signal.}
    \label{fig:noise_comp}
\end{figure}
\begin{figure}[t]
    \centering
    \includegraphics[width=0.45\textwidth]{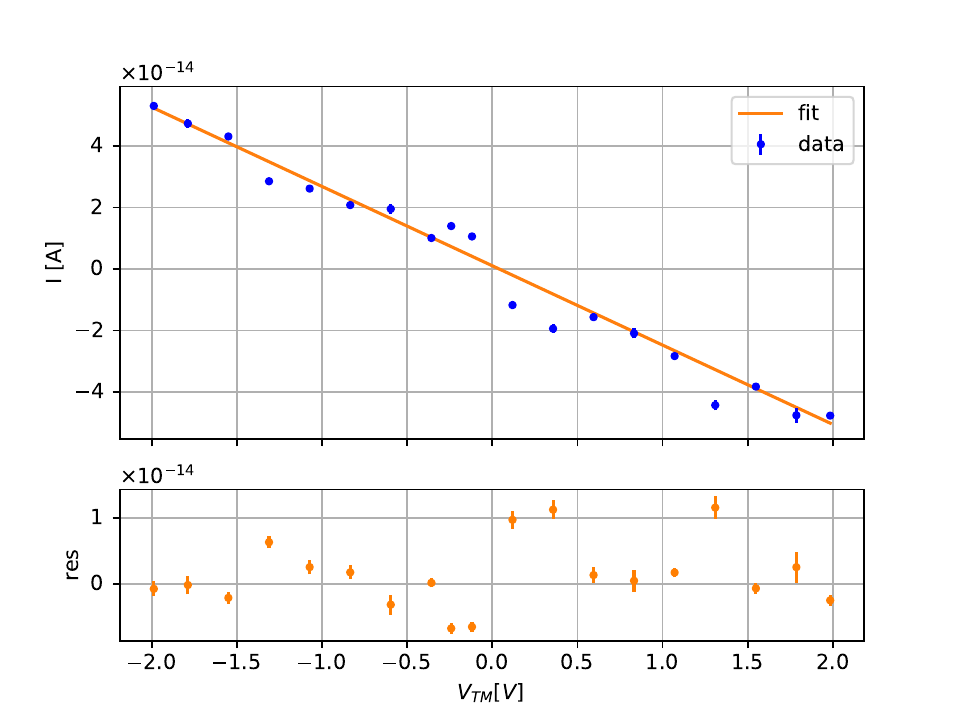}
    \caption{Dark current of the BART apparatus measured as angular coefficient of the i-V curve measured without UV illumination.}\label{fig:dark_current_fit}
\end{figure}
\begin{figure*}
    \centering
    \includegraphics[width=0.45\linewidth]{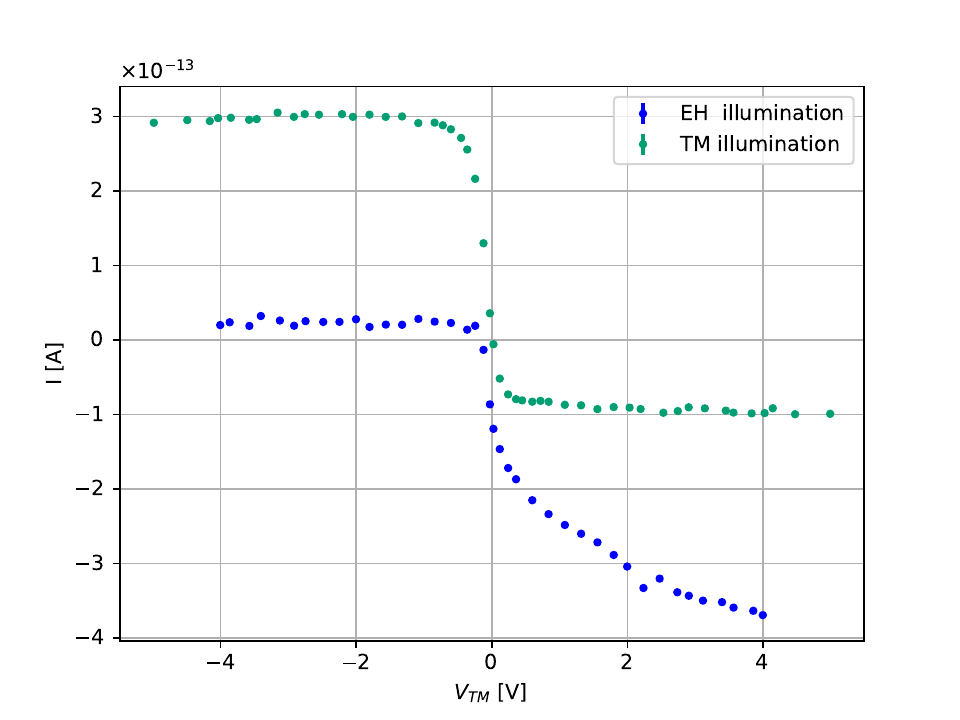}
    \includegraphics[width=0.45\textwidth]{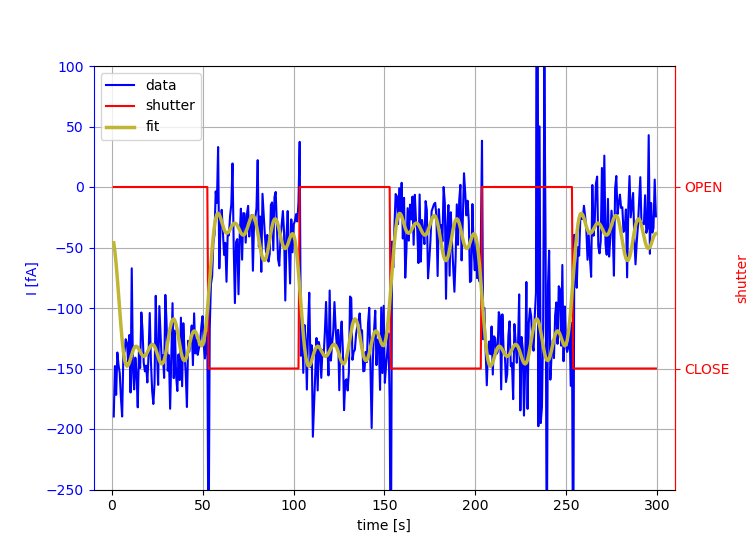}
    \caption{Left: Discharge curves for test mass and electrode housing illumination schemes. Error bars are contained within the point. Right: Example of a single photo-current measurement and fit, obtained with EH UV illumination and $V_{TM}=1.8\,\text{V}$.}\label{fig:UV_discharges}
\end{figure*}

\section{Performance testing of BART} \label{sec_comm}

To ensure the correctness electronic configuration and that pressure level was low enough in the vacuum chamber, a series of preliminary tests had been carried out. 

First, noise measurements of the electrometer were made while keeping $V_{TM}=0\,V$. The electrometer was left in acquisition for two consecutive nights, one at full scale range of 20 pA, used during UV discharge and the lowest energy beam measurements ($<100\,$M$e$V), the other one at at the 200 pA full scale used at the higher proton energies. From the spectra in Figure \ref{fig:noise_comp}, we estimate resolution of $\sigma_{20\;pA} = 0.4\;\text{fA}$ and $\sigma_{200\;pA} = 1.2\;\text{fA}$ measuring for $5$ minutes a $10\;\text{mHz}$ signal.  These are more than enough for the expected beam produced current of the order of $1-60\,$pA.

Proceeding to the experimental validation of the apparatus, an initial verification of the correct electrical configuration is performed by measuring the so‑called “dark current”, i.e. stray current due to not perfect isolation of the TM from EH, such as the dashed line in Figure \ref{fig:scheme2}. Figure \ref{fig:dark_current_fit} shows the current measured at different $V_{TM}$ and a fit to Ohm's law. We find $R_{leak} = 38 \pm 1\;\text{T}\Omega$ compatible with typical nominal Teflon resistance.  
This value of $R_{leak}$ induces detectable current offsets (of the order of $100\,$fA - 1 pA) by our apparatus, motivating the need for modulation of the current sources as described below. 

As introduced before, in LISA, the TM will be kept neutral within a few mV via the photo-electric effect under UV illumination. This produces a photo-current in the 10-500 fA range (known from other experiments and LPF) made  $\lesssim$ $1$\,eV electrons. Since this is 1-2 orders of magnitude lower than the expected pA generated by the proton beam, the demonstration of the ability of the BART apparatus to measure it is a good test of the limit sensitivity of the apparatus and of the correct electrical configuration.

Figure \ref{fig:UV_discharges}  shows the photo-current measured through the electrometer for different values of the TM potential, while illuminating either TM and EH with the UV LED.

These ``discharge curves'' follow the typical behavior observed in LPF and in the torsion pendulums in Trento \citep{PhysRevD.107.062007_Charging2023,dalbosco2023,chiavegato2024}. Both curves exhibit a saturation behavior when $|V_{TM}|\geq 1$\,V. This behavior is expected as the residual kinetic energy of the emitted photoelectrons is $K_e = E_{UV} - W_{AU} < 1$\,$e$V. At negative $V_{TM}$ electrons are accelerated away from the TM and attracted by the EH. Conversely at positive $V_{TM}$, they are attracted by the TM and repelled by the EH. The EH curve exhibits a less sharp saturation for positive TM due to some light shining partially in the gaps between electrodes and EH.

These measurements are performed modulating the light source by a 10\,mHz in-fiber shutter (chopper) while keeping the test mass at a fixed $V_{TM}$ for 3 periods  (5 minutes). Figure  \ref{fig:UV_discharges} shows an example of the measured current superimposed to the shutter status and the square-wave fit. Odd harmonics Fourier series with an offset is fit to the measured data
$$s(t) = a_0 + \sum_{i_{odd}}^{9} A_n \cos(\omega_n t) + B_n \sin(\omega_n t)$$.
For each $V_{TM}$ the fit is repeated over three periods, and the resulting parameters are averaged to estimate the statistical uncertainty. The first harmonic component gives the photo-current estimate as $i_0 = (\pi/4) B_n$. This procedure, which allowed for achieving a resolution of $10\,\text{fA}$, was used to measure the proton beam induced charging current. This measurement scheme is the basis for the ``end-to-end'' discharge test in preparation for the LISA Engineering Model of the GRS \citep{Venturelli_2026}.

\section{Experimental Procedure}\label{sec_proc}
\begin{figure*}
    \centering
    \includegraphics[width=0.4\linewidth]{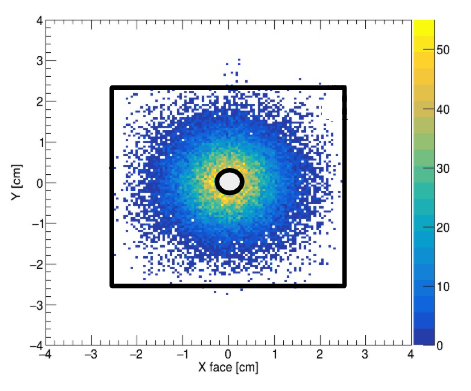}
    \includegraphics[width=0.4\linewidth]{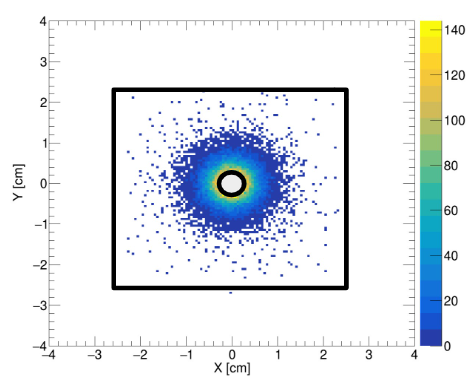}
    \caption{Distribution of primary protons from 72 M$e$V (left) and 228 M$e$V (right) beams on the GRS face (black solid contour) simulated by TMCTK. The center of the GRS face features a 6 mm diameter hole, as described in Section \ref{sec_sim}. Coloured axis scale refers to number of simulated events.}
    \label{fig:simface}
\end{figure*}

\begin{figure*}
    \centering
    \includegraphics[width=0.45\linewidth]{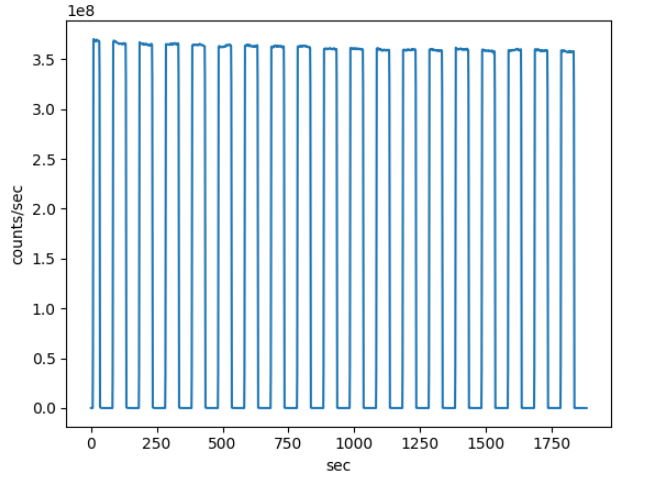}
    \includegraphics[width=0.45\linewidth]{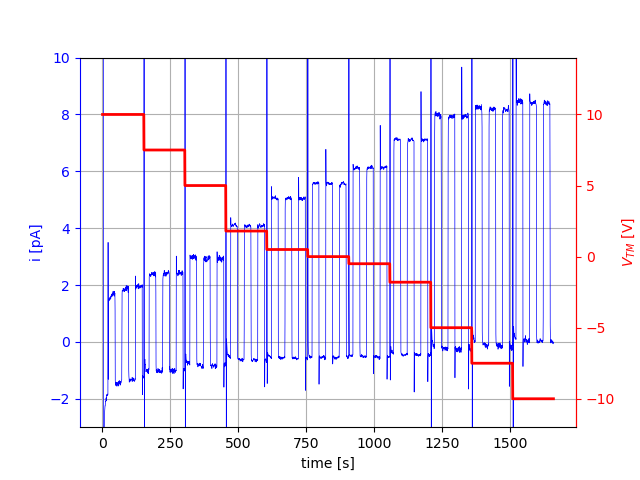}
    \caption{Example of time series acquired during irradiation at 154~M$e$V. 
Left: proton counts measured by the proton-therapy (PT) ionization chamber. 
Right: charging current measured by the BART apparatus; the lower panel shows the time dependence of the bias voltage applied to the electrode housing (EH) during the measurement.}
    \label{fig:squarebeam}
\end{figure*}

\begin{table*}
    \centering

    \begin{tabular}{ccccc}
        \hline
        Beam energy [M$e$V]&
        Flux on target [p/s] &
        Beam FWHM [mm] &
        0-bias current on face X [pA] \\
        \hline\hline
        72.0 $\pm$ 0.7& (5.50 $\pm$ 0.27) $\times$ 10$^7$ & 15.9 &0.561 $\pm$ 0.003 \\
        83.0 $\pm$ 0.8& (6.75 $\pm$ 0.33) $\times$ 10$^7$& 15.2&0.430 $\pm$ 0.004\\
        91.0 $\pm$ 0.9& (1.00 $\pm$ 0.05) $\times$ 10$^7$& 14.6&1.711 $\pm$ 0.005\\
        98.0 $\pm$ 1.0& (1.20 $\pm$ 0.06) $\times$ 10$^8$& 13.7& 6.50 $\pm$ 0.02\\
        108.0 $\pm$ 1.0& (1.05 $\pm$ 0.05) $\times$ 10$^8$& 13.4&16.43 $\pm$ 0.02\\
        125.0 $\pm$ 1.2& (1.65 $\pm$ 0.08) $\times$ 10$^8$& 12.3& 23.79 $\pm$ 0.04\\
        136.0 $\pm$ 1.4& (1.15 $\pm$ 0.60) $\times$ 10$^8$ & 11.3&16.70 $\pm$ 0.01\\
        154.0 $\pm$ 1.5& (0.95 $\pm$ 0.05) $\times$ 10$^8$& 10.2& 13.60 $\pm$ 0.01\\
        198.0 $\pm$ 2.0& (3.70 $\pm$ 0.18) $\times$ 10$^8$& 8.1& 42.25 $\pm$ 0.06\\
        228.0 $\pm$ 2.3& (9.00 $\pm$ 0.45) $\times$ 10$^8$& 6.8& -2.249 $\pm$ 0.008\\
        \hline
    \end{tabular}

    \vspace{5mm}


    \begin{tabular}{ccccc}
        \hline
        Beam energy [M$e$V]&
        Flux on target [p/s] &
        Beam FWHM [mm] &
        0-bias current on face Y [pA] \\
        \hline\hline
          72.0 $\pm$ 0.7& (5.20 $\pm$ 0.26) $\times$ 10$^7$& 15.9 & 0.384 $\pm$ 0.003\\
        83.0 $\pm$ 0.8& (6.50 $\pm$ 0.32) $\times$ 10$^7$& 15.2&1.188 $\pm$ 0.002\\
        91.0 $\pm$ 0.9& (1.02 $\pm$ 0.05) $\times$ 10$^8$& 14.6&3.562 $\pm$ 0.006\\
        108.0 $\pm$ 1.0& (1.10 $\pm$ 0.05) $\times$ 10$^8$& 13.4& 17.44 $\pm$ 0.03\\
         136.0 $\pm$ 1.4& (1.20 $\pm$ 0.06) $\times$ 10$^8$& 11.3&17.22 $\pm$ 0.02\\
        154.0 $\pm$ 1.5&(1.00 $\pm$ 0.05) $\times$ 10$^8$ & 10.2&14.15 $\pm$ 0.02\\
        198.0 $\pm$ 2.0& (3.85 $\pm$ 0.19) $\times$ 10$^8$& 8.1& 42.9 $\pm$ 0.1\\
        228.0 $\pm$ 2.3& (1.00 $\pm$ 0.05) $\times$ 10$^9$& 6.8& 84.21 $\pm$ 0.09\\
        \hline
    \end{tabular}

    \caption{Charging current measured by the BART experiment at different beam energies (X face at top, Y face bottom). The data refers to measurements done with V$_{TM}-$V$_{EH}$=0. The table reports also the beam measured flux on target and the nominal beam width at the impact point, i.e. BART vacuum chamber.}
    \label{tab:beam}
\end{table*}

The experimental room of the PT facility shown in Figure \ref{fig:exp_room} is equipped with a laser alignment system that allows the precision positioning of the apparatus under test on the three axes, marking the exact point in space where the proton beam exhibits the nominal intensity and width quoted for every available beam energy. During the assembly of the BART apparatus, we marked on the external of the vacuum chamber the positions corresponding to the centers of the X,Y sensor faces. Using this reference, we were able to align the proton beam toward the center of the TM with an estimated accuracy of 2 mm. The final distribution of protons hitting the GRS face under study depends, besides on the precision of the alignment, on the Full Width at Half Maximum (FWHM) of the proton beam. In Table \ref{tab:beam} the beam energy used in this experiment and their corresponding beam FWHM and proton flux intensities are reported. Typical distribution of protons hitting the GRS, obtained using the TMCTK simulation, are shown in Figure \ref{fig:simface} for two different beam energies. The impact of the  hole (i.e. number of protons going through the hole and thus not passing through the EH material) is estimated varying from $\sim$ 10\% to $\sim$ 20\% passing from lower energies, where the beam is less focused, to higher ones. The proton beam intensities,  were chosen such as to guarantee an expected measured TM charging current of the order of 1-100 pA, easily measurable by the electrometer. 

As done in the performance testing phase (Section \ref{sec_comm}), we employed 10 mHz square wave modulation of the proton irradiation at every energy in order to disentangle the measurement of the charging current from any possible current bias or drift. During the irradiation, we varied in steps the EH potential in the [-10,10] V range or in the [-2,2] V range, with the TM held at virtual ground by the electrometer input. We measured three periods of the modulation square wave for each potential, for a total irradiation time of the order of 30 minutes per beam energy. As an example, the left panel of Figure \ref{fig:squarebeam} shows the proton flux time series of one particular BART measurement campaign, corresponding to beam energy of 154 M$e$V, measured by the ionization chamber of PT facility. The right panel of the same Figure shows the corresponding charging current measured by the BART apparatus, for different values of bias potential V$_{EH}$ applied to the EH every three cycles of the modulating square wave. As can be seen, the measured current value is fully time correlated with the on/off status of the beam, confirming the detection of the TM charging process.
Again, the amplitude of the charging current was extracted through Fourier analysis of the current time series measured by the electrometer (see Section \ref{sec_comm}).

\section{Discussion of results}
In this Section, we present the BART experiment results: we first discuss the measurements of TM charging with V$_{TM}-$V$_{EH}$=0 (Section \ref{zerobias}). Next, we present the variation of the charging current as a function of the TM-EH potential difference (Section \ref{sec_observation}) and compare it with (Section \ref{sec_simcomp}) the expectations from the modified TMCTK toolkit described in Section \ref{sec_sim}. From the observed variation of charging current, we explore for the first time the energy distribution of the LEE (Section \ref{spectrum}).

\subsection{Net charging with 0V bias} \label{zerobias}
\begin{figure}[t]
\centering
\includegraphics[width=0.51\textwidth]{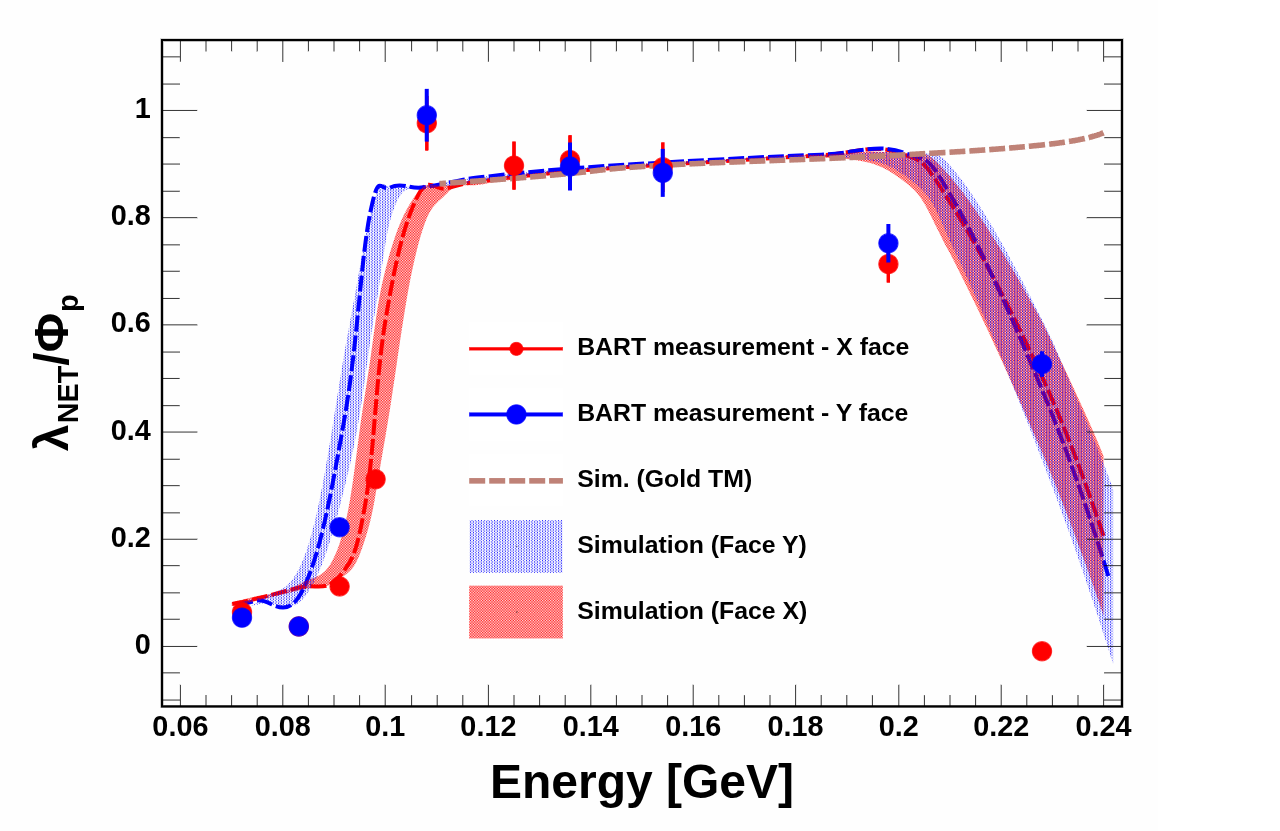}
\caption{Net charging per incident proton ($\lambda_p/\Phi_P$) measured with the BART experiment irradiating X (red points) and Y (green points) of the GRS, as a function of proton beam energy. The experimental data refers to measurements done with V$_{TM}-$V$_{EH}$=0. Continuous lines, dashed lines and shaded areas represent the corresponding results from TMCTK simulations, as described in the text.}\label{fig:lambda_net}
\end{figure}

The net charging of the TM with no bias applied to EH is shown in Figure \ref{fig:lambda_net} in comparison to TMCTK simulation results. The quantity in the Y axis is the net charging $\lambda_{NET}$ normalized by the intensity of the incident proton flux $\Phi_p$. In the simulation, $\lambda_{NET}$ is computed \citep{ARAUJO2005451} as:
$$\lambda_{NET}=\sum_{j=-\infty}^{j=+\infty}j \lambda_j,$$ where $j$ is the net signed charge deposit in units of elementary charge e on the TM and $\lambda_j$ is its rate of occurrence measured in s$^{-1}$. The equivalent rate intensity $\Phi_p$ is computed starting from the number of simulated proton events N, as $\Phi_p = N/\Delta t$, where $\Delta t$ is the equivalent simulated beam time. The resulting quantity $\lambda_{NET}/\Phi_p$ is dimensionless and is a measurement of the average contribution of every proton at different energies to the total net charging of the TM in units of e.

The result obtained from the simulation is shown as dashed lines in Figure \ref{fig:lambda_net}. Green (red) line represents the result obtained with the simulation tuned to match face X (Y) material distribution. The shadowed uncertainty bars include uncertainty from material budget estimation and from beam energy spread.

 Starting from left (lower beam energies) one can observe the energy region below 90 M$e$V where protons are unable to penetrate the material around the TM, resulting in a negligible TM charging. Next, a sharp rise between 90 and 100 M$e$V marks the energy threshold where the protons start getting through the surrounding material and hit the TM. After the threshold, the trend shows a plateau with $\lambda_{NET}/\Phi_p$ value between 0.8 and 0.9. The value, close but smaller than one, shows that the dominating phenomenon here is the stopping of the protons in the TM material, charging it with a +$e$. The plateau value ($<$1) is compatible with previous TMCTK results with CR-like irradiation \citep{toolkit}, which showed slightly higher electron emission from the EH than from the TM. This causes more electrons to enter the TM with respect to those leaving it, yielding an overall negative average contribution of electrons to TM charging in this energy range. The plateau continues until the proton energy becomes sufficient for complete penetration of the TM, where a drop to $\lambda_{NET} \sim $ 0 is expected. In the flight model, characterized by Pt-Au TM, this drop-off is expected at energies of several G$e$Vs. In the BART experiment, exploiting the less dense copper bulk of the TM, this effect is observed in the PT beam energy range, marking at $\sim$180 M$e$V the limit of the representativeness of this test for the flight conditions. \textcolor{blue}{We note that the directional irradiation used in BART does not reproduce the approximately isotropic cosmic-ray environment expected in flight. Under isotropic irradiation, and in the presence of the non-uniform spacecraft mass distribution surrounding the GRS, the sharp charging threshold observed in Fig.~\ref{fig:lambda_net} is expected to be smoothed, and the absolute charging rate may differ at the level of the geometrical uncertainties typically observed in charging simulations. However, the characteristic energy scale associated with the onset of net charging is expected to remain largely unchanged. The results in Fig.~\ref{fig:lambda_net} should therefore be interpreted as a benchmark of the beamed BART configuration and of the low-energy part of the charging process, rather than as a direct prediction of the in-flight LISA charging spectrum.}

\begin{figure*}
    \centering
    \includegraphics[width=0.4\linewidth]{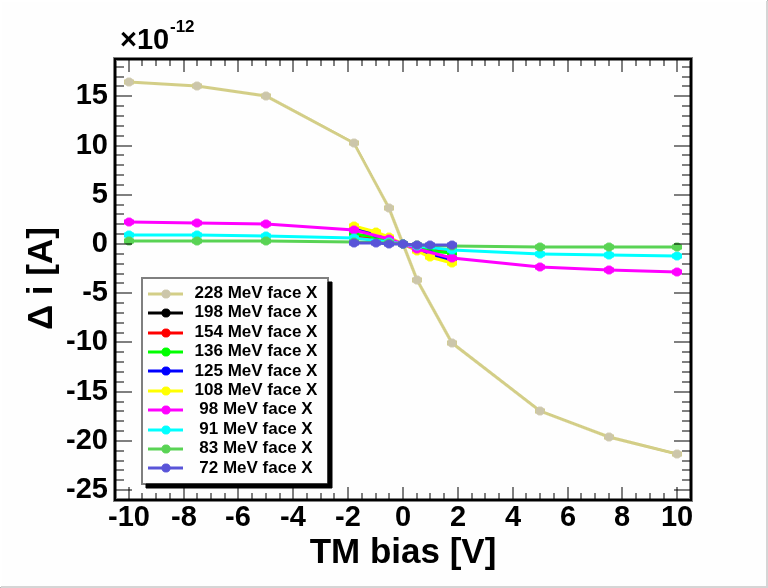}
    \includegraphics[width=0.4\linewidth]{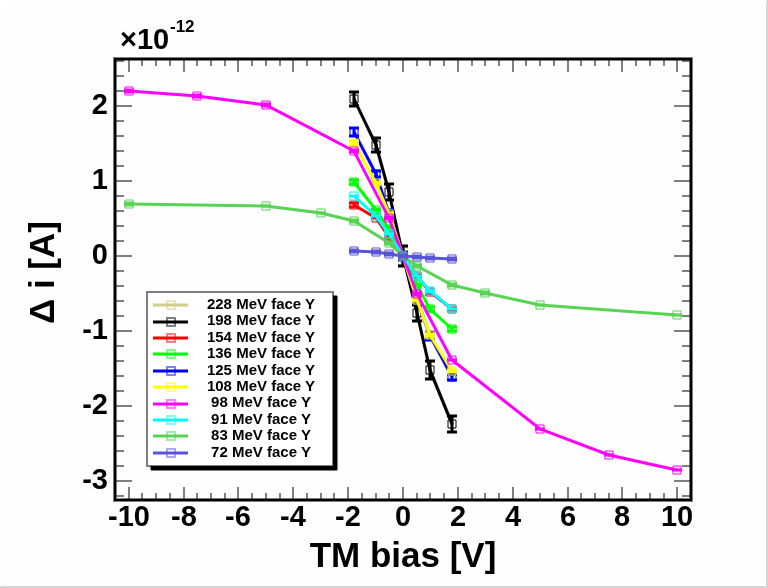}
    \includegraphics[width=0.42\linewidth]{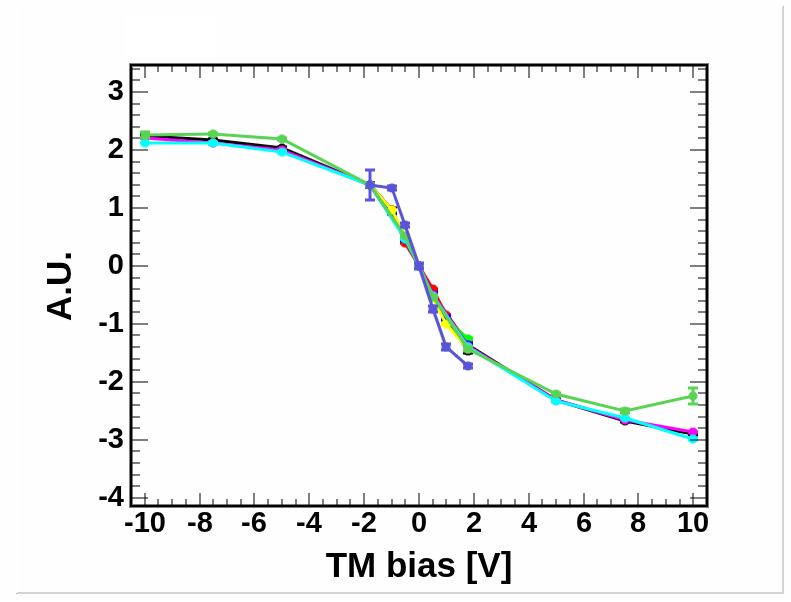}
    \includegraphics[width=0.41\linewidth]{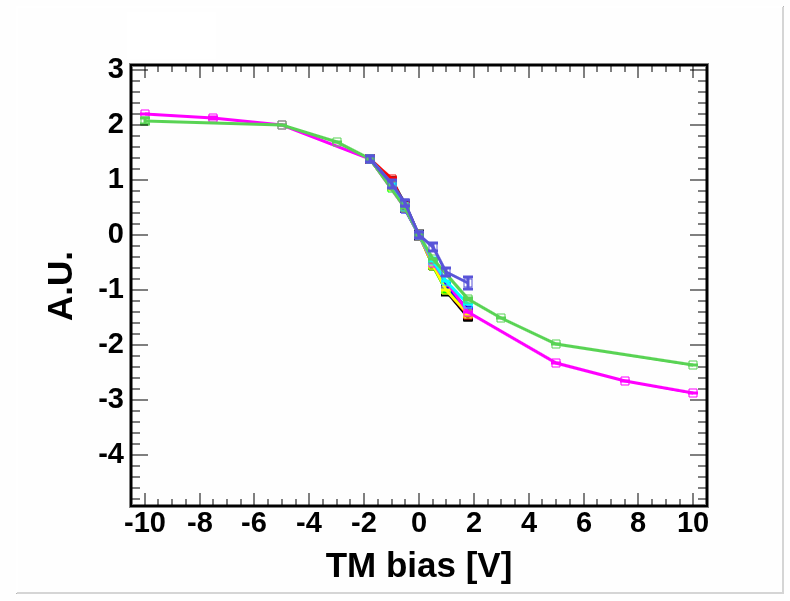}
    \caption{Top: Measurements of charging current variations $\Delta i$ with respect to the current measured at 0 bias for different beam energies for X (left) a Y (right) face irradiation, as a function of TM voltage bias. Bottom: Same measurements are shown normalized to superimpose on each other (in Arbitrary Units).}
    \label{fig:currentvar}
\end{figure*}

The experimental data for $\lambda_{NET}$ were calculated as 
$$\lambda_{NET}=i_{ch}/e,$$ where $i_{ch}$ is the measured value of the charging current extracted using the Fourier square wave analysis described in Section \ref{sec_proc} and $e$ is the elementary charge, while $\Phi_p$ can be directly measured using the PT ionization chamber, put in front of the apparatus, obtaining an estimation of $\lambda_{NET}/\Phi_p$. The uncertainty on the quantity is dominated by the 5\% uncertainty quoted for the ionization chamber measurement of $\Phi_p$. The uncertainty on measured currents, obtained from the fit procedure explained in Section \ref{sec_comm}, is reported in Table \ref{tab:beam} and results negligible with respect to the ionization chamber measurement one, thanks to the low noise level ($\sim$ 10 fA/$\sqrt{\mbox{Hz}}$) of the electrometer. Other sources of uncertainty in the current measurement not perfect beam alignment and pointing towards the center of the GRS face. From dedicated studies done with the TMCTK they resulted $<$1\% in the whole energy range, and were added in quadrature the ones already described. The results regarding the measured $\lambda_{NET}/\Phi_p$ are also reported in Figure \ref{fig:lambda_net}. The two sets of measurements regards two different measurement campaigns, obtained irradiating the faces X and Y of the sensor. As can be seen, the overall qualitative behavior of both measurements is in line with the expectations from the TMCTK simulations, showing the expected sharp rise at proton energies between 90 and 100 M$e$V and a gradual decrease after 170 M$e$V. Below 140 M$e$V, the level of difference between the two sets of measurement are compatible with the spread of simulation results caused by the grammage difference of the two faces of the EH. Moreover, as expected, the denser face X shows a threshold at slightly higher energies with respect to the Y face. The experimental data show instead a significant difference with respect to the simulation at high energy, between 180 and 240 M$e$V: the decrease in data is more gradual, and starts at lower energies. At 228 M$e$V the TM charging drops to some small negative values while irradiating the X face, suggesting that all the protons crossed the TM without depositing charge. The difference with respect to simulation and more importantly to Y face illumination was unexpected and also contradictory to the fact that for Y-face beaming  that behavior did not occur\footnote{ Face Y provides $\sim$ 1 g/cm$^2$ less grammage than face X}. A dedicated investigation of this discrepancy was conducted using both numerical simulations and experimental measurements. A first study was performed to disentangle a possible effect of the holes in faces X and Y, systematically varying the beam impact point on the GRS surface by up to 1 cm. Moreover, on the simulation side, an additional analysis was performed to quantify the effect of uncertainties in beam alignment and TM density. None of the variability induced by these effects was sufficient to account for the observed discrepancy.
A possible explanation may involve a surplus of electron emission from surfaces in this energy range, in particular the secondary population of electrons predicted by the simulations at energies between 30 and 100 $e$V (see Figure \ref{fig:lee}).
\textcolor{blue}{Such electrons cannot be directly investigated with the present setup because the maximum voltage bias that can be applied between TM and EH is 10 V, limiting the reconstructed electron spectrum to energies below approximately 10 eV}. Additional experimental studies are therefore required to test this hypothesis, to be addressed in future work through a systematic exploration over a broader range of bias potentials.

\subsection{Observations of LEE in charging with variable TM - EH voltage bias}\label{sec_observation}

In this section we report the results obtained varying the voltage bias on the EH.

Based on the results of the TMCTK \citep{toolkit}, the energy distribution of the LEE emitted from the gold surfaces is strongly peaked at kinetic energies below $\sim$4 $e$V, so the strongest variation of charging is expected at very low voltage biases. For this reason, for most of the beam energies reported in Table \ref{tab:beam} we performed a scan in voltage in the range [-2,2] V with a granularity of 7 different voltages. For several values of energies (91 and 98 M$e$V for face Y, 83, 91, 98 and 228 M$e$V for face X) we explored a wider range, [-10,10] V, to search for a saturation effect possibly caused by the sharp drop of the LEE energy spectrum at energies above $\sim$ 5 $e$V observed in the simulations \citep{toolkit}.

\begin{figure*}
    \centering
    \includegraphics[width=\linewidth]{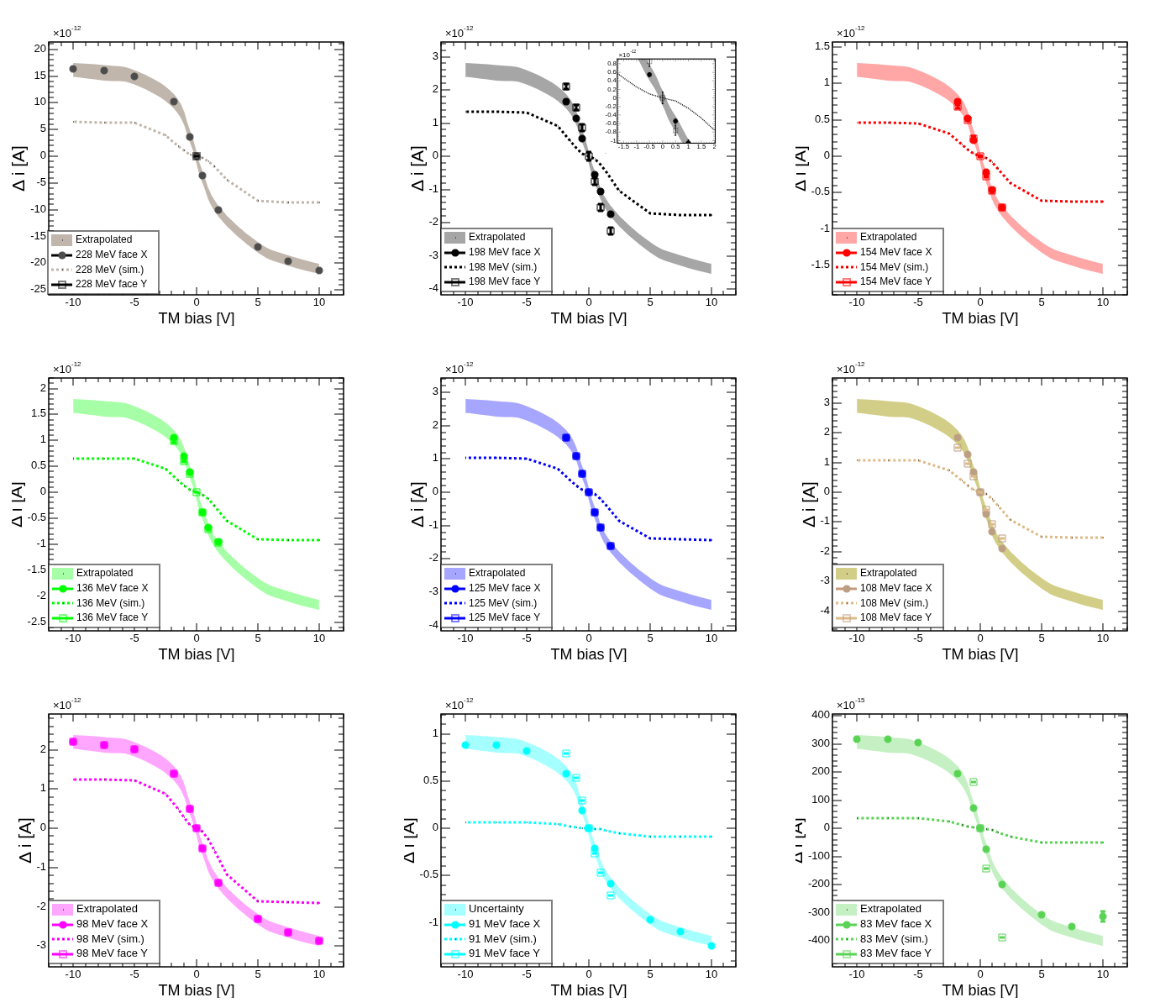}
   \caption{Comparison of the measured and simulated $\Delta i$ as a function of the applied test mass bias voltage. 
BART measurements are shown for face~X (filled squares) and face~Y (open squares) and are compared with the prediction of the joint TMCTK--COMSOL simulation tool (dashed line). 
The shaded region indicates the extrapolation of the face~X measurement over the voltage range $[-10,\,10]$~V. The insert in the top center panel shows a zoom of the center part of the plot.}
    \label{fig:curr_comparison}
\end{figure*}

In the two top panels of the Figure \ref{fig:currentvar} the results of this investigation are shown, reported as variation of measured current 
$\Delta i \left( V \right) \equiv i \left( V \right) - i \left( 0 \right)$, where $i  \left( 0 \right)$ is the charging current measured with 0 V bias applied (reported in Table \ref{tab:beam}), as a function of the TM potential bias with respect to EH. As can be seen, the current variation due to redirection of LEE is significantly visible for all all beam energies. Negative TM biases lead to a positive variation of charging current, consistent with negative TM repelling LEEs that would otherwise have hit its surface. The opposite is true for positive TM bias. This is the first experimental evidence of the role of LEE in a controlled ground experiment with a flight-representative GRS sensor. 
As can be seen, the curves show different $\Delta i$ amplitudes, indicating that the amount of LEE  changes with beam energy. The shape of the curves results instead qualitatively compatible between each beam energy. This is shown in the lower panels of  Figure \ref{fig:currentvar}, where all the experimental curves were normalized to superimpose on each other: clearly the energy spectrum of LEE is poorly dependent on the kinetic energy of the primary particle,
as expected for a phenomenon purely linked to the work function of the TM and EH surfaces. We exploit this feature  to extrapolate those measurements done in the  $\pm$ 2V  EH bias $V_{EH}$ interval to $\pm$ 10 V and compare them with the TMCTK simulations.
Finally, the measurements done in the wider voltage range show signs of a charging current saturation for voltage biases higher in modulus than 4 V, especially on the negative bias sector, indicating that the majority ($>$ 80\%) of LEE have a kinetic energy below $\sim$ 4 eV.

\subsection{LEE emission: comparison with simulations}\label{sec_simcomp}

In this Section the results discussed in Section \ref{sec_observation} are compared to the results obtained from the with the COMSOL electrostatic model of the TMCTK toolkit, as described in Section~\ref{sec_sim}. Exploiting the similarity of the measured $\Delta i$ curves, the experimental results are extrapolated to the [-10,10] V bias voltage range for the cases in which only the measurement in the [-2,2] V range was performed, using the curve obtained from 98 M$e$V on Y face measurement as a template. The measurements for X and Y face and the extrapolations are shown in Figure \ref{fig:curr_comparison}, compared with simulation results. As can be seen, the results from simulations show a qualitatively similar trend to the measured ones, exhibiting a similar saturation behavior for TM bias voltages above 4 V. 
 The experimental and simulated curves exhibit a different shape near the origin, as shown in the insert in the top center panel of Figure \ref{fig:curr_comparison}. That has been linked to the modeling of the angular dependence of the gold work function discussed in Section \ref{sec_sim}. In particular, the almost null derivative of the simulated curve around the 0 V bias suggests that the combined effect of simplified geometry and emission angular dependence in the model penalizes too much the emission at lower electron energies (below $\sim$ 1 eV). A deeper investigation, both on the experimental and the simulation side is needed to confirm this insight. 

 The $\Delta$i saturation value can be considered in both measurements and simulations as a proxy of the total size of the pool of LEE escaping the surfaces and migrating from TM and EH. As can be seen from Figure \ref{fig:curr_comparison}, this magnitude is systematically underestimated by the current simulations, by a factor of 2-3. 

\begin{figure}[t]
\centering
\includegraphics[width=0.5\textwidth]{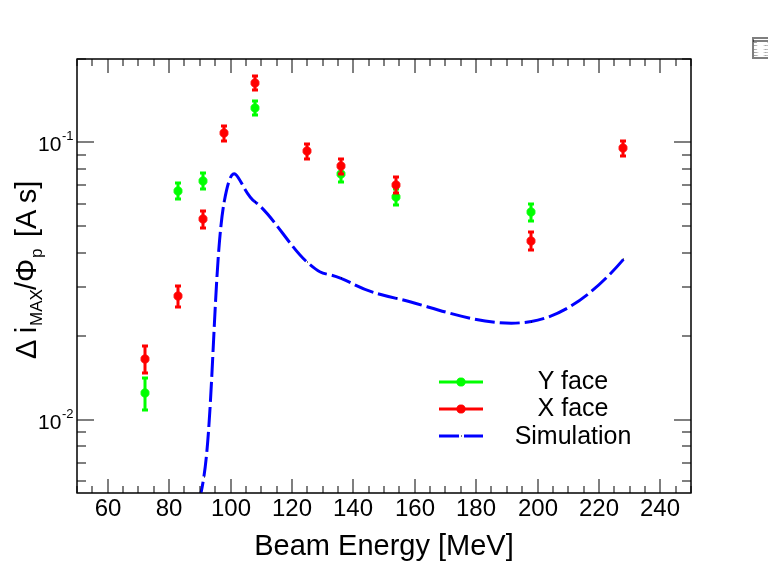}
\caption{Value of $\Delta$i at -10 V TM bias voltage, normalized by incident proton flux, as a function of proton beam energy measured from BART irradiating face X (red) and Y (green), compared with results from COMSOL-TMCTK simulation tool (blue dashed).  }\label{fig:saturation}
\end{figure}

To further characterize LEE emission, we studied the negative saturation value of $\Delta i$ as a function of the proton beam energy, to estimate the relative contribution of LEE to the total TM charging. The results, normalized by the incident proton flux, are shown in Figure ~\ref{fig:saturation} for all measured beam energies.  The experimental data show that LEE are visible from primary proton energies above the  $\sim$80--90~M$e$V threshold\footnote{required for protons to enter the EH-TM gap and thus cross the EH and TM gold plated surfaces}, peaking at about 110 M$e$V. We note that data from X and Y face beaming are slightly shifted in energy, with the latter anticipating the LEE production consistently with a lower entrance threshold for the protons.

As the proton energy increases further, the smaller amount of ionization leads to a decrease in LEE production. Above $\sim$200 M$e$V, an increasing fraction of primary protons have sufficient energy to fully traverse the TM, enabling additional LEE emission from the rear surfaces of both the TM and the EH. This effect produces a second increase in the measured $|\Delta i|/\Phi_p$.

The same qualitative behavior is reproduced by the simulations with a lower intensity, but with a steeper activation threshold around 100 M$e$V. 
This difference is plausibly again attributable to the simplified geometric representation adopted in the simulation, which assumes a uniform EH thickness across the GRS face. In contrast, the real sensor features a local variability in the material distribution due to the presence of the gaps separating the electrodes from the grounded housing, allowing a fraction of protons with energies below $\sim$80~M$e$V to reach the TM. This results in the more gradual onset observed in the experimental data.

\begin{figure*}
    \centering
    \includegraphics[width=0.7\linewidth]{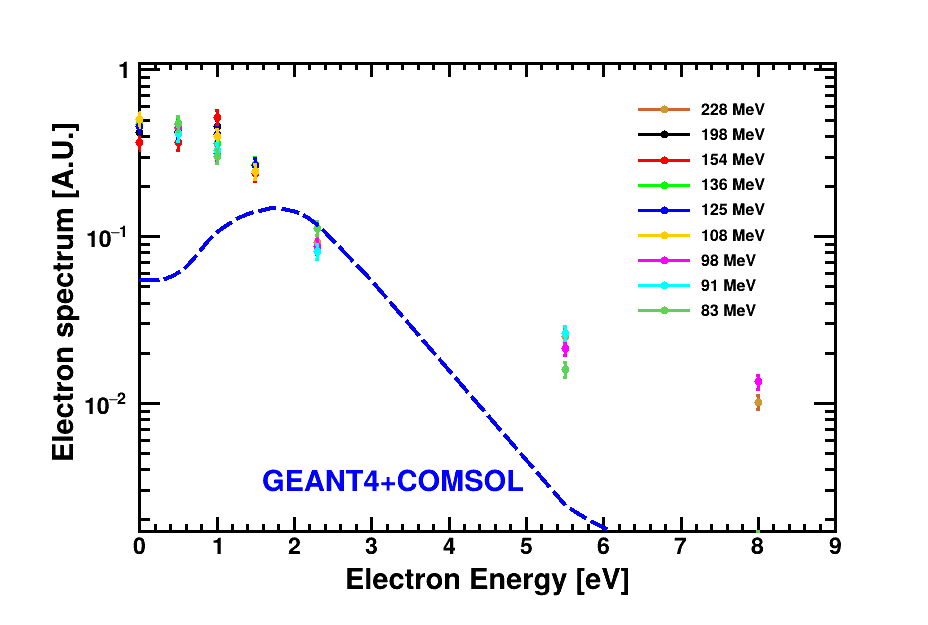}
    \caption{Measurements of LEE energy spectra emitted in the TM-EH gap measured by BART, compared with the prediction of the TMCTK simulation (GEANT4 module + COMSOL electrostatic FEM for LEE propagation in TM-EH gap) described in Section \ref{sec_sim}.}
    \label{fig:spectraLEE}
\end{figure*}

\subsection{Characterization of LEE emission spectrum} \label{spectrum}
Starting from the experimental curves shown in Figure~\ref{fig:curr_comparison}, it is possible to extract information on the energy distribution of the LEE emitted from the gold surfaces of the TM and EH. In particular, we focus on the negative-saturation region of the $\Delta i(V)$ curves. The applied -V negative TM bias acts as an electrostatic barrier for electrons emitted from the EH. Electrons with kinetic energy below $e|$V$|$ are prevented from reaching the TM, while only those with higher energy contribute to the measured charging current. As a result, increasing the magnitude of the negative TM bias progressively suppresses electrons of higher and higher kinetic energy.

The variation of the measured current between two consecutive bias values, $V_{j-1}$ and $V_j$, therefore is directly proportional to the number of electrons with with kinetic energies in the interval
\[
e|V_j| < E_e < e|V_{j-1}|.
\]
This provides a direct way to probe the energy distribution of the emitted LEE. In practice, we can estimate the  differential energy spectrum of the EH emitted LEE from the derivative of the negative-saturation branch of the $\Delta i(V)$" curve as a numerical derivative of the negative-saturation branch of the $\Delta i(V)$ curve:
\[
\delta_j = \left.\frac{\delta(\Delta i)}{\delta V}\right|_{V_j},
\]
where $V_j$ is the $j^{\mathrm{th}}$ applied TM bias voltage. The quantity $\delta_j$ is proportional to the number of electrons in the corresponding energy interval and can therefore be interpreted as an approximated estimate of the differential energy spectrum of the emitted LEE, neglecting higher-order contributions, including those arising from the geometry of the GRS and from the distribution of electron emission angles.

Figure~\ref{fig:spectraLEE} shows the normalized quantity $\delta_j / i_{\max}$ as a function of the electron kinetic energy, approximated as $E_e \simeq e|V_j|$. Here, $i_{\max}$ denotes the charging current measured at (or extrapolated to) a TM bias of $-10$~V, corresponding to the regime in which essentially all EH emitted electrons are suppressed, while TM emitted electrons are unaffected. The normalization allows comparison of the spectral shape across different beam energies. The measured spectra are compatible with each other, that is the density of states of LEE is not dependent on the energy of the primary. In the same Figure, the same calculation performed on the curves obtained through the joint TMCTK-COMSOL simulation model at a beam energy of 137 M$e$V is shown as a comparison. Spectra obtained from simulations at other beam energies are very similar to this one. The comparison shows that experimental data suggest a much flatter spectrum for LEE with respect to the prediction of the model, and an important relative excess of LEE at energies below 2 $e$V and above 5 $e$V.
\textcolor{blue}{ The discrepancy between simulated and measured spectra may partly originate from the modeling of electron emission from gold surfaces: While the low-energy region is sensitive to the modeling of the work function (and possible surface contamination), the harder tail observed above about 4 eV is largely insensitive to it, indicating limitations in the current emission model.}

\section{Conclusions}
The charging of free-falling test masses by energetic particles is a significant disturbance source for space-based gravitational-wave observatories. Although numerical simulations have been extensively employed to model this process, their validation has been constrained by the scarcity of controlled experimental data—most notably regarding the influence of low-energy secondary electrons. In this work, we present and implement an accelerator-based experimental campaign specifically designed to directly measure proton-induced test mass charging in a configuration representative of the LISA Gravitational Reference Sensor.

The experiment demonstrates that irradiating a flight-representative test mass and electrode housing with mono-energetic proton beams in the 70–230 M$e$V energy range yields measurable charging currents at the pico-ampere level, substantially above  electrometer system’s instrumental noise floor, confirming the suitability of the experimental apparatus for precision charging investigations. The measured charging rates are consistent in magnitude with expectations derived from prior simulation studies, once scaled to the applied beam intensities.

A central experimental finding is the pronounced dependence of the measured charging current on the applied test mass potential. Systematic variations of the test mass bias over a range of order 10 V produce reproducible changes in the net charging rate, indicating the presence of charge carriers whose trajectories and collection efficiencies are strongly influenced by the local electrostatic potential configuration. This behavior provides direct experimental evidence that low-energy secondary electrons emitted from internal surfaces make a significant contribution to the overall charging dynamics.

To our knowledge, these measurements constitute the first controlled laboratory observation of proton-induced test mass charging in a LISA-like sensor geometry, with explicit sensitivity to low-energy electron processes occurring within the narrow gap between the test mass and the electrode housing. The results confirm the physical picture inferred from LPF observations, in which secondary electron emission plays a non-negligible role in determining both the magnitude of the charging response and its dependence on the electrostatic boundary conditions.

Another important experimental result of this work is the first measurement of the energy distribution of proton-induced low-energy electrons emitted in a representative geometry of the LISA Gravitational Reference Sensor. The spectrum was reconstructed from the dependence of the charging current on the applied test mass bias, which acts as an electrostatic energy filter.

The measured LEE spectrum was found to be largely independent of the energy of the incident protons over the explored range. A comparison with predictions obtained using a state-of-the-art GEANT4-based test mass charging simulation toolkit—employing the GEANT4-DNA library to model LEE production at gold surfaces and COMSOL to simulate their propagation in the electrostatic fields of the sensor gap—reveals significant discrepancies. In particular, the measurements show a pronounced excess of electrons at kinetic energies below 2~eV and a harder spectral component above 5~eV. Integrated over the 0--10~eV range, the total low energy electron yield exceeds the simulation expectations by a factor of approximately 2--3.

These results provide new experimental constraints on low-energy electron emission processes in LISA-like sensors. They highlight limitations in the current modeling of surface emission and electron transport and represent an essential benchmark for the validation and further development of simulation tools used to predict test mass charging in space-based gravitational-wave observatories.
Moreover, they show that accelerator-based experiments can directly probe charge-deposition and secondary-emission mechanisms under well-controlled conditions, reducing reliance on simulations alone. More broadly, this work provides a practical, scalable framework for validating key disturbance mechanisms for next-generation precision space interferometry missions.

\section*{Acknowledgement}
The authors thank Prof. Giacomo Ciani (University of Trento), Prof. Stefano Vitale (University of Trento) and Dr. Francesco Nozzoli (INFN-TIFPA) for helpful discussions about the experimental setup and the interpretation of results. 

This work was funded by ASI – Agenzia Spaziale Italiana – accordo attuativo no. 2024-36-HH.0 dell’ accordo quadro ASI/Università di Trento no. 2017-32-H.0 – Addendum no. 2017-29-H.1-2020 All’Accordo no. 2017-29-H.0. The authors acknowledge support from the Istituto Nazionale di Fisica Nucleare (INFN) through the CSN2 research program.

\bibliographystyle{unsrt}
\bibliography{aipsamp}
\end{document}